\documentclass[pre,aps,twocolumn,amsmath,amssymb]{revtex4-1}

\pdfoutput=1
\usepackage{hyperref}
\usepackage{graphicx}
\usepackage[usenames]{color}
\usepackage{bm}
\usepackage[final]{movie15}

\def\eqq#1{Eq.~(\ref{#1})}
\def\eq#1{(\ref{#1})}
\def\av#1{\langle #1 \rangle}

\def\f#1{Fig.~\ref{#1}}

\def\c#1{~\cite{#1}}

\def\x{{\bm x}}

\def\s#1{Section~\ref{#1}}

\def\e{{\rm e}}

\def\beq{\begin{equation}}
\def\eeq{\end{equation}}
\def\bea{\begin{eqnarray}}
\def\eea{\end{eqnarray}}

\begin{document}

\title{Multi-point nonequilibrium umbrella sampling and associated fluctuation relations}
\author{Stephen Whitelam}
\email{{\tt swhitelam@lbl.gov}}
\affiliation{Molecular Foundry, Lawrence Berkeley National Laboratory, 1 Cyclotron Road, Berkeley, CA 94720, USA}
\begin{abstract}
We describe a simple method of umbrella trajectory sampling for Markov chains. The method allows the estimation of large-deviation rate functions, for path-extensive dynamic observables, for an arbitrary number of models within a certain family. The general relationship between probability distributions of dynamic observables of members of this family is an extended fluctuation relation. When the dynamic observable is chosen to be entropy production, members of this family include the forward Markov chain and its time reverse, whose probability distributions are related by the expected simple fluctuation relation. 
\end{abstract}

\maketitle

\section{Introduction} 

One concept widely used to quantify rare behavior in computer simulations is importance sampling, in which the typical behavior of a reference system is used to infer the rare behavior of a model of interest\c{frenkel2001understanding}. Umbrella sampling\c{torrie1977nonphysical} is a method of importance sampling used in equilibrium. Here, a model is modified by a bias potential that constrains it to a region of interest, and knowledge of the Boltzmann weights of the modified and original models allows the user to calculate thermodynamic properties of the latter. Away from equilibrium there exist methods similar in spirit to umbrella sampling. These methods probe the statistical properties of dynamic trajectories, using a reference dynamics whose typical behavior is in some sense equivalent to the rare behavior of the model of interest\c{bucklew1990large,touchette2009large,garrahan2009first, giardina2006direct,maes2008canonical,giardina2011simulating,lecomte2007numerical,nemoto2014computation,rohwer2015convergence,nemoto2016population,ray2017importance,jpgdimer,ray2017exact}. 
\begin{figure*}[t] 
   \includegraphics[width=0.8\linewidth]{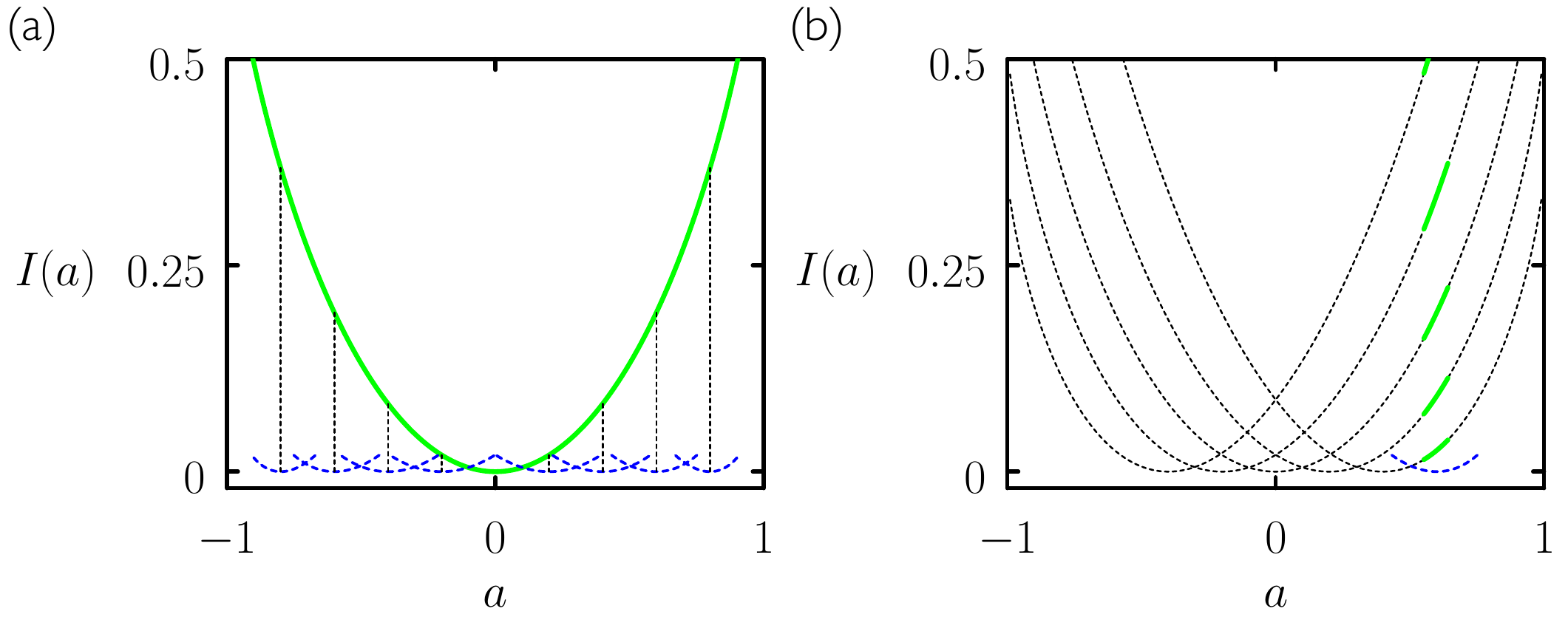} 
   \caption{Nonequilibrium umbrella sampling. (a) Estimation of the rate function $I(a)$ of a model $s$ (green), a member of the family \eq{rm}, can be done using a set of reference models $s'$. Typical reference-model trajectories, i.e. trajectories whose values of $a$ concentrate near the minima $a_{s'}$ of the reference-model rate functions (blue dashed lines), each allow reconstruction of one point $I(a_{s'})$ on the green curve. Unlike conventional umbrella sampling, each piece of $I(a_{s'})$ is evaluated independently, and values of $s'$ can be as close or as far apart as desired. (b) Multi-point sampling. In this paper we show that each set of reference-model simulations can be used to recover one point (middle of the green interval) on the rate functions of any set $\{s\}$ of models of the family \eq{rm}. Thus a collection of reference-model simulations allows reconstruction of the rate functions of the entire family of models.}
   \label{fig0}
\end{figure*}

One such method, described in Refs.\c{klymko2017rare,lattice2}, applies to Markov chains of a fixed number of configuration changes. The method makes use of a set of reference models to reconstruct the large-deviation rate functions, for path-extensive dynamical observables, for an original model of interest. The rate functions of the reference models can be regarded as nonequilibrium umbrella potentials, concentrating sampling in particular regions of parameter space, an idea sketched in \f{fig0}(a). The concept of trajectory importance sampling is well known\c{bucklew1990large,touchette2009large,garrahan2009first, giardina2006direct,maes2008canonical,giardina2011simulating,lecomte2007numerical,nemoto2014computation,rohwer2015convergence,nemoto2016population,ray2017importance,jpgdimer,ray2017exact}; the approach of Refs.\c{klymko2017rare,lattice2}  is different to most approaches in that it uses as a reference dynamics a simple (rather than an optimal) modification of the original dynamics, and does not use population dynamics (a.k.a. cloning) methods\c{giardina2011simulating}. 

In general, trajectory importance sampling of the probability distribution $\rho(a) \sim \e^{-K I(a)}$ of a path-extensive quantity $K a$ requires evaluation of relations of the type
\beq
\label{e_one}
\e^{-K I(a)} \sim \e^{s K a} \int {\rm d} q \, P_s(q) e^{K q}.
\eeq
Here $I(a)$, the large-deviation rate function for $a$ for the original model, is the quantity we want to calculate; $s$ is a parameter associated with the reference model; $q$ is a fluctuating quantity associated with trajectories of the reference model; and $K$, the trajectory length, is a large parameter. $P_s(q)$ is the distribution of $q$ associated with an ensemble of reference-model trajectories. The formally optimal approach to trajectory sampling determines (or approximates) the reference model for which $P_s(q) = \delta(q)$, so that reference-model trajectories of given $a$ are a constant factor less probable than those of the original model\c{chetrite2013nonequilibrium,jack2015effective,chetrite2015variational,garrahan2016classical,ferre2018adaptive}. In this case the effort of the method lies in the construction of a (potentially complex) reference model whose properties are ideal in the stated sense. The present method uses instead a reference-model set that is a simple modification of the existing model, and the effort of the method focuses on evaluation of $P_s(q)$ in order to recover $I(a)$. 

Direct evaluation of \eq{e_one} is impractical when $K$ is large. If $P_s(q)$ is Gaussian (the best case scenario outside of $q$ being constant) with variance $\sigma^2$, then the error in the measurement of $e^{K q}$ over trajectories of the reference model is
\beq
\label{e_var}
\frac{\av{(e^{K q})^2}_s-\av{e^{K q}}_s^2}{\av{e^{K q}}_s^2} \sim \e^{K^2 \sigma^2},
\eeq
where $\av{\cdot}_s \equiv \int {\rm d} q\, (\cdot) P_s(q)$. The variance $\sigma^2$ is typically $\propto 1/K$, in which case \eq{e_var} diverges exponentially with $K$, and a number of trajectories exponentially large in $K$ is needed to evaluate \eq{e_one}. However, it is not necessary to evaluate \eq{e_one} or measure $e^{K q}$ directly (indeed, for $K$ of even modest size such exponentials cannot be evaluated on a computer). Taking logarithms of \eq{e_one} gives
\beq
\label{e_two}
I(a) = -s a - \bar{q} - K^{-1}\ln \int {\rm d} q \, P_s(q) \e^{K (q-\bar{q})},
\eeq
where $\bar{q}$ is the mean value of $q$ over the ensemble of reference-model trajectories. The first two terms of \eq{e_two} can calculated straightforwardly, using a number of trajectories that does not scale with $K$ (see Appendix C of\c{klymko2017rare}), and provide a upper bound on the rate function $I(a)$. Any choice of reference dynamics produces {\em some} upper bound on the rate function\c{varadhan2010large}; we show that certain simple and physically-motivated choices produce meaningful (i.e. tight) bounds. Moreover, with some additional numerical effort the third term in \eq{e_two} can be evaluated or closely approximated, yielding the rate function itself. For the case of Gaussian $P_s(q)$, the third term is equal to $K\sigma^2/2$, which can be evaluated by direct simulation using a number of trajectories that scales less quickly than linearly with $K$ (see Appendix C of\c{klymko2017rare}). Higher-order cumulants of the integral become progressively more expensive to evaluate -- and the method will fail if $P_s(q)$ has long tails that cannot be sampled efficiently by direct simulation -- but in what follows we show that a set of simple reference-model choices can yield short-tailed $P_s(q)$, in which case the rare behavior of the original model can be evaluated by direct simulation of a set of simple reference models.

Thus the method described in Refs.\c{klymko2017rare,lattice2} is a particular implementation of trajectory importance sampling, and is non-optimal in the formal sense. However, it is simple and it works: it requires no special techniques beyond direct simulation, and can be used to bound or closely approximate the rate function in a variety of settings. In addition, the method directly computes the logarithmic probability distribution for path-extensive observables, rather than computing the Legendre transform of the distribution. This feature is useful when the probability distribution has linear or non-convex portions, in which case its Legendre transform is singular\c{touchette2009large}. 

In this paper we describe two extensions of this method. First, we point out that each set of reference dynamical trajectories can be used to generate pieces of the probability distribution of {\em any} stochastic model within a particular family. We refer to this procedure as {\em multi-point} sampling, an idea sketched in \f{fig0}(b). Second, we extend the method to Markov chains of fixed time. 

The paper is organized as follows. In \s{sampling} we describe the multi-point sampling procedure. We show that the general relationship between probability distributions of dynamic observables of any two members of this family is an extended fluctuation relation. The sampling method proceeds by evaluating many distribution-pairs at the concentration point of one distribution, so allowing evaluation of the other distribution away from its concentration point. In \s{ent} we show that when the dynamical observable is entropy production, models in this family include forward and time-reversed trajectory ensembles, whose distributions satisfy the expected fluctuation relation\c{gallavotti1995dynamical,kurchan1998fluctuation,maes1999fluctuation,lebowitz1999gallavotti,crooks2000path,seifert2005entropy,speck2012large}. In \s{cts} we show how to carry out the sampling method for trajectories of fixed time. We conclude in \s{conc}.

\section{Importance sampling of dynamic trajectories} 
\label{sampling}

The reference-model method of Refs.\c{klymko2017rare,lattice2} is designed to calculate probability distributions of dynamic observables for Markov chains of a fixed number of configuration changes. Consider a family of models, parameterized by a variable $s$, that move between microstates $C$ and $C'$ with probability 
\beq
\label{pb}
p_s(C \to C') = \frac{W_s(C \to C')}{R_s(C)}.
\eeq 
Here 
\beq
\label{rm}
W_s(C \to C') = {\rm  e}^{-s \alpha(C \to C')} W_0(C \to C')
\eeq
is a rate, and $R_s(C) \equiv\sum_{C'} W_s (C \to C')$ is a normalization factor than ensures probability conservation, i.e. $\sum_{C'} p_s(C \to C')=1$. The quantity $\alpha(C \to C')$ is the change of an arbitrary dynamic observable $A$ upon moving from $C$ to $C'$. The form \eq{rm} is motivated by the exponential tilting of the probability distribution commonly done in the mathematical literature\c{dembo2010large,touchette2009large}, although this choice of rates does not in general produce the tilted distribution.

Our aim is to calculate the probability distribution 
\beq
\label{pdf}
\rho_s(a,K) \equiv \sum_\x{P_s[\x] \delta{(A[\x]-K a)}} 
\eeq
of the intensive variable $a=A/K$, for a particular model $s$, over trajectories $\x = \{C_0,C_1,\dots,C_K\}$ of $K$ steps of the dynamics. (Other approaches to trajectory sampling calculate the cumulant-generating functions of dynamic observables\c{lebowitz1999gallavotti,garrahan2009first,pietzonka2016universal}.) In \eq{pdf}, the quantity
\beq
P_s[\x] =\rho_s(C_0) \prod_{k=0}^{K-1} p_s(C_k \to C_{k+1})
\eeq
is the probability of generating a trajectory $\x$ within the model $s$, and
\beq
A[\x] = \sum_{k=0}^{K-1} \alpha(C_k \to C_{k+1})
\eeq
is the extensive dynamic observable for the trajectory $\x$. Direct simulation of the model leads to good sampling of $\rho_s(a,K)$ for $a \approx a_s$, the value of $a$ typical of the model. For models with a well-defined stationary measure
\beq
\label{stat}
\pi_s(C) = \sum_{C'} \pi_s(C') p_s(C' \to C),
\eeq
this value is given by
\bea
a_s&=&\sum_C \pi_s(C) \sum_{C'} p_s(C \to C') \alpha(C \to C') \nonumber \\
&=&-\sum_C \pi_s(C)\partial_s \ln R_s(C).
\eea
\begin{figure*}[t] 
   \includegraphics[width=\linewidth]{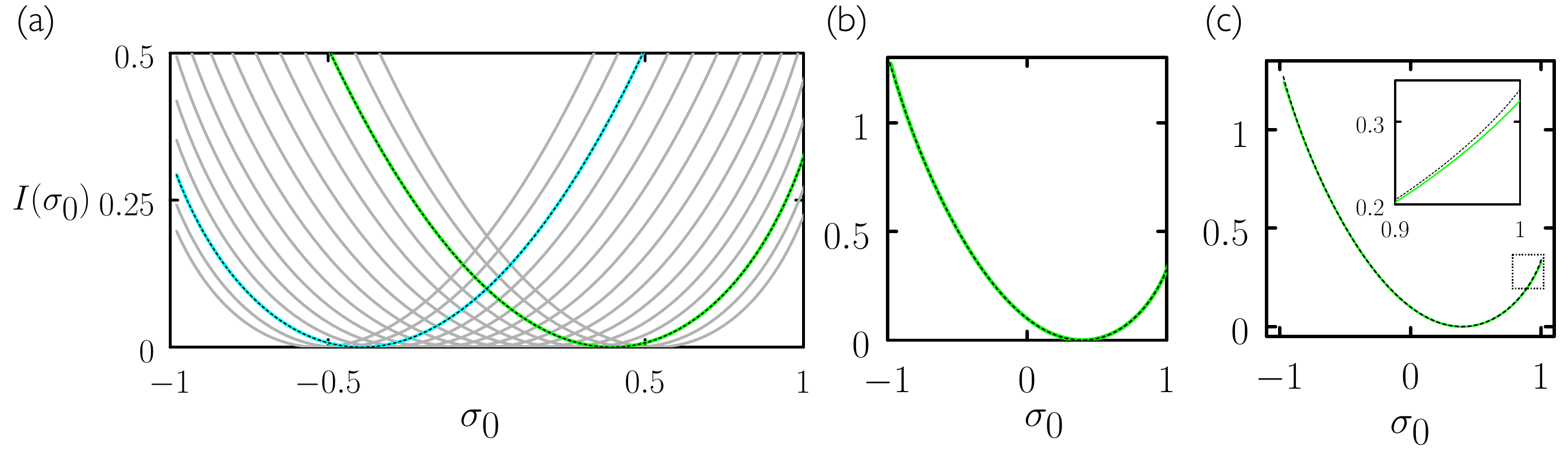} 
   \caption{(a) Large-deviation rate functions for entropy production, $I(\sigma_0) = -K^{-1} \ln \rho(\sigma_0,K)$, for 15 models $s=\{-1.2,1.1,\dots,0.1,0.2\}$ (left to right) in the family \eq{rm}. The `original' model ($s=0$) is the the 4-state model of Ref.\c{gingrich2016dissipation}. Results were calculated using the multi-point nonequilibrium umbrella sampling scheme described in the text. The green and cyan curves denote the original ($s=0$) and reverse ($s=1$) models; the black dashed lines, which denote $I_0(\sigma_0)+\sigma_0$ on the left and $I_1(\sigma_0)-\sigma_0$ on the right, verify that these two models obey the fluctuation relation \eq{oneb}. Note that entropy production $\sigma_0$ is defined with respect to the model $s=0$. (b) Rate function for the model $s=0$ computed using the present method (green) and by diagonalization of the tilted transition matrix (black). (c) Comparison of the exact result (green) and the rate-function upper bound, \eqq{three} (black). Inset: zoom-in of the dotted area.}  
   \label{fig1}
\end{figure*}
(The method also works for models that do not have a well-defined stationary measure\c{klymko2017rare}.)

Trajectories of the model will concentrate on $a_s$ in the large-$K$ limit, leading to good sampling there but poor sampling elsewhere. A standard way to overcome this problem is to use importance sampling\c{bucklew1990large,touchette2009large,garrahan2009first,maes2008canonical,giardina2011simulating}. Consider a second model $s'$, called the reference model, whose trajectories concentrate in the large-$K$ limit on $a_{s'}$. Direct simulation of the reference model will lead to good sampling in the vicinity of $a_{s'}$, and this statistics can be used to reproduce the piece $\rho_s(a_{s'},K)$ of the original model's distribution. To see how, note that the relative likelihood with which a trajectory $\x$ is generated by reference and original models, $w_{ss'}[\x]=P_s[\x]/P_{s'}[\x]$, is 
\beq
w_{ss'}[\x] =e^{(s'-s) A[\x]+ K q_{ss'}[\x]},
\eeq
where
\beq
\label{kew}
q_{ss'}[\x] \equiv K^{-1} \sum_{k=0}^{K-1} \ln \frac{R_{s'}(C_k)}{R_s(C_k)}.
\eeq
The quantity we want, the probability distribution of $a=A/K$ for trajectories of length $K$ of the original model, is
\bea
\label{com}
\rho_s(a,K)&\equiv& \sum_\x{P_{s'}[\x] w_{ss'}[\x] \delta{(A[\x]-K a)}} \nonumber\\
&=& \e^{(s'-s) a K} \sum_\x{P_{s'}[\x] \e^{K q_{ss'}[\x]} \delta{(A[\x]-K a)}} \nonumber\\
&\equiv& \rho_{s'}(a,K) \e^{(s'-s) a K} \av{\e^{K q_{ss'}[\x]}}_{s'}^{a K},
\eea
where
\beq
\av{\cdot }_{s'}^{a K} \equiv  \frac{\sum_\x{P_{s'}[\x] (\cdot) \delta{(A[\x]-K a)}}}{\sum_\x{P_{s'}[\x] \delta{(A[\x]-K a)}}}
\eeq
is an average over reference-model trajectories that possess $A[\x]= Ka$. \eqq{com} can be written 
\beq
\label{fluc}
\frac{\rho_s(a ,K)}{\rho_{s'}(a ,K)} = \e^{(s'-s) a K} \int {\rm d} q_{ss'} \, \e^{K q_{ss'}} P_{s'}(q_{ss'}|a),
\eeq
where $P_{s'}(q_{ss'}|a)$ is the probability distribution of the fluctuating piece of the path weight, $q_{ss'}[\x]$, for trajectories of the reference model $s'$ that possess $A[\x]=K a$. Normalization is such that $\int {\rm d} q_{ss'}\, P_{s'}(q_{ss'}|a)=1$. For models for which the stationary measure $\pi_s(C)$ exists, values of $q_{ss'}[\x]$ fluctuate, from trajectory to trajectory, about the typical value
\beq
\label{pw}
q_{ss'} = \sum_C \pi_{s'}(C) \ln \frac{R_{s'}(C)}{R_s(C)}.
\eeq
\eqq{fluc} involves no approximations: it is a re-writing of \eq{pdf}, and is valid for arbitrary path length $K$. The expression makes clear that the trajectory-ensemble distributions of $a$ for two models, $s$ and $s'$, are related by an extended fluctuation relation\c{touchette2009large}. 

This relation can be used as a tool for sampling the distribution of a model $s$ using a reference model $s'$. Consider the case of large $K$. For many models, for sufficiently large $K$, the distribution $\rho_s(a, K)$ adopts the large-deviation form $\rho_s(a, K)\sim \e^{-K I_s(a)}$, where $I_s(a)$ is the large-deviation rate function. To use \eq{fluc} as a method of reconstructing $I_s(a)$, note that the equation can be evaluated at the point $a=a_{s'}$ where trajectories of the reference model concentrate, i.e. where $I_{s'}(a_{s'})=0$. Then \eq{fluc} can be rewritten
\bea
\label{one}
I_s(a_{s'}) &=& (s-s') a_{s'} -\bar{q}_{ss'}\nonumber \\
&-&K^{-1} \ln \int {\rm d} q_{ss'} \, \e^{K (q_{ss'}-\bar{q}_{ss'})} P_{s'}(q_{ss'}|a_{s'}), \hspace{0.4cm}
\eea
whose evaluation gives one point on the curve $I_s(a) = -K^{-1} \ln \rho(a,K)$. (We omit the limit $K \to \infty$ in the definition of the large-deviation rate function, in order to emphasize that we calculate quantities for large but finite values of path length. We have verified that the precise value of $K$ makes no difference to the plots in the paper, provided that it is large enough.) Here $\bar{q}_{ss'}$ is the mean of the quantity $q_{ss'}[\x]$, for trajectories of the reference model $s'$ that have $A[\x] = K a_{s'}$. The variance of the integral in \eq{one} diverges (in general) exponentially with $K$, but its logarithm can, in the cases we have encountered\c{klymko2017rare,lattice2}, be evaluated without excessive numerical effort. Upon simulating the reference model, we can measure $P_{s'}(q_{ss'}|a_{s'})$ (provided that this quantity does not have long tails that cannot be sampled efficiently by direct simulation). If this distribution happens to be Gaussian in $q_{ss'}$, then \eq{one} can be evaluated analytically to give
\beq
\label{two}
I^{(1)}_s(a_{s'}) = (s-s') a_{s'} -\bar{q}_{ss'}-K \sigma^2_{ss'}/2.
\eeq
For non-Gaussian weight fluctuations we can regard \eqq{two} as an approximation of the rate function. Here $\sigma^2_{ss'}$ is the variance of the quantity $q_{ss'}[\x]$, for trajectories of the reference model $s'$ that have $A[\x] = K a_{s'}$. In this case the numerical effort required to evaluate the point $I_s(a_{s'})$ is (only) the effort required to compute the variance (from trajectory to trajectory) $\sigma^2_{ss'} \propto 1/K$ of the intensive quantity $q_{ss'}[\x]$. If $P_{s'}$ is non-Gaussian in $q_{ss'}$ then the integral in \eq{one} can be evaluated via calculation of higher-order cumulants, with the associated additional computational cost\c{klymko2017rare}. The fact that the integral in \eq{one} is exponentially large in $K q[\x]$ does not automatically defeat the re-weighting procedure: if the distribution $P_{s'}(q_{ss'}|a_{s'})$ can be sampled efficiently by the reference model dynamics (e.g. if it is approximately Gaussian) then \eq{two} can be evaluated with reasonable accuracy.
~\\
\indent The first two terms on the right-hand side of \eqq{one} provide, by Jensen's inequality, an upper bound $I^{(0)}_s(a_{s'})$ on the rate function piece $I_s(a_{s'})$ (whether or not $P_{s'}$ is Gaussian),
~\\
\beq
\label{three}
I_s^{(0)}(a_{s'}) = (s-s') a_{s'} -\bar{q}_{ss'}.
\eeq
~\\
This bound, like the rate function itself, is not required to be quadratic about its minimum or even to be convex\c{klymko2017rare,lattice2}. It can be calculated using a single reference-model trajectory, or analytically if $\pi_s(C)$ can be calculated analytically\c{lattice2}.
~\\
\indent In Refs.\c{klymko2017rare,lattice1,lattice2} we used this method to calculate large-deviation rate functions for a single model $s=0$ using a set of reference models $s'$, a procedure illustrated in \f{fig0}(a). The rate functions of the reference models serve as nonequilibrium umbrella potentials, concentrating sampling at certain points in parameter space. In this paper we point out that Equations \eq{fluc} and \eq{one}, evaluated using a reference model $s'$, can be used to produce large-deviation rate functions for any number of `original' models $\{s\}$, without doing additional simulations. The quantity $q_{ss'}[\x]$, given by \eqq{kew}, depends on the identity of the original model only through the label $s$; the choice of microstates visited is determined solely by the reference model $s'$. Thus by keeping track, in a reference-model simulation, of the set of $N$ values $q_{\{s\} s'}[\x]$, where $\{s\} =\{s_1,\dots,s_N\}$, we can recover $N$ pieces $I_{\{s\}}(a_{s'})$ of $N$ distinct `original' models. This idea is illustrated in \f{fig0}(b). Then, by scanning $s'$, we can simultaneously reconstruct $N$ different rate functions $I_{\{s\}}(a)$.
~\\
\section{Sampling entropy production} 
\label{ent}
To illustrate this procedure we choose to sample entropy production. Let the bias appearing in \eqq{rm} be
\beq
\label{alpha}
\alpha(C \to C') =\ln \frac{\pi_0(C) p_0(C \to C')}{\pi_0(C')p_0(C' \to C)},
\eeq
the entropy produced by the model $s=0$ upon moving from $C \to C'$; here $\pi_0(C)$ is the stationary measure of the model $s=0$ [see \eqq{stat}]. The extensive dynamical order parameter is then the total entropy produced by a trajectory $\x$ of $K$ steps, $A[\x] = \sum_k \alpha(C_k \to C_{k+1}) \equiv \Sigma_0[\x]$. The rate of entropy production is the intensive version of this parameter, $\sigma_0[\x] = \Sigma_0[\x]/K$. In this case the family of models \eq{rm} is 
\bea
\label{rmep}
W_s(C \to C') &=& \left( \frac{\pi_0(C')p_0(C' \to C)}{\pi_0(C)p_0(C \to C')} \right)^s W_0(C \to C'). \hspace{0.9cm}
\eea
The parameter $s$ influences the rate of entropy production (here defined with respect to the model $s=0$), distributions of which obey the expression \eq{fluc}. 

The distributions of certain members of this family are related in a simple way. In particular, the model $s=1$ is the reverse of the model $s=0$. From \eq{rmep} we have
\beq
W_1(C \to C') = \frac{\pi_0(C')}{\pi_0(C)} R_0(C) p_0(C' \to C).
\eeq
Then 
\beq
\label{r1}
R_1(C) \equiv \sum_{C'} W_1(C \to C') = R_0(C),
\eeq
using \eq{stat}, and so 
\beq
p_1(C \to C') = \frac{\pi_0(C')}{\pi_0(C)} p_0(C' \to C),
\eeq
which is the time-reversed Markov chain\c{crooks2000path}. The probability distributions of entropy produced by the models $s=0$ and $s=1$ then obey the expected fluctuation relation. Setting $s=0$ and $s'=1$, the terms of the fluctuating piece $q_{ss'}[\x']$ of the path weight, \eqq{kew}, are
\beq
\ln \frac{R_1(C)}{R_0(C)} = 0,
\eeq
using \eq{r1}, giving $q_{01}[\x]=0$ and $P_1(q_{01}|a) = \delta(q_{01})$. Hence for $s=0$ and $s'=1$, \eqq{fluc} reduces to
\beq
\label{oneb}
\frac{\rho_0(\sigma_0,K)}{\rho_1(\sigma_0,K)} = \e^{\sigma_0 K}.
\eeq 
Given that the entropy production $\sigma$ changes sign if we switch from the forward to the reverse model, i.e. $\sigma_0 = -\sigma_1$, we can write \eq{oneb} in the form 
\beq
\label{fr}
\frac{\rho(\Sigma)}{\rho_{\rm rev}(-\Sigma)} = \e^{\Sigma},
\eeq 
which is the fluctuation theorem for nonequilibrium steady states\c{gallavotti1995dynamical, kurchan1998fluctuation,maes1999fluctuation,lebowitz1999gallavotti,seifert2005entropy}. Here `rev' denotes the reverse model, and $\Sigma = \sigma K$. Thus, the general expression \eq{fluc} relates the distribution of an arbitrary dynamic observable $A$ for any two members of the the family of models $s$. If $A$ is chosen to be entropy production $\Sigma$, then, for $0 < s < 1$, these models interpolate between the forward and reverse versions of the model whose rates are $W_0(C \to C')$.

We can use these fluctuation relations, in the form \eq{one}, as a method for reconstructing the large-deviation rate functions for entropy production for all models $\{s\}$ in the family \eq{rm}. To illustrate this procedure we consider the 4-state model of Ref.\c{gingrich2016dissipation}. This is a fully connected model with states $C \in \{1,2,3,4\}$. We take the rates $W_0(C \to C')$ to be the quantities $r(C,C')$ in the caption of Fig. 2 of that paper:
\bea
\begin{array}{lll} 
    r(1, 2) = 3,& r(1, 3) = 10,& r(1, 4) = 9, \nonumber\\
    r(2, 1) =10,& r(2, 3) = 1,& r(2, 4) = 2,\nonumber\\ 
    r(3, 1) = 6,& r(3, 2) = 4,& r(3, 4) =1,\nonumber \\
r(4, 1) = 7,& r(4, 2) = 9,& r(4, 3) = 5,\\
   \end{array}
   \eea
with $r(C,C)=0$. We carried out a series of simulations of 100 different reference models, whose values of $s'$ run from $-2$ to $2.5$ in evenly-spaced intervals. For each reference model $s'$ we generated $8 \times 10^3$ independent trajectories of length $K=10^6$. From this ensemble we calculated the typical value $a_{s'}$ of the dynamical observable, and identified those trajectories whose values of $a$ were within an interval $\pm10^{-3}$ of $a_{s'}$. From this restricted ensemble we collected values of the fluctuating path weights $q_{\{s\},s'}[\x]$, for 15 different values $\{s\}$, evenly spaced between $-0.2$ and $1.2$. 
 \begin{figure*}[] 
    \centering
    \includegraphics[width=\linewidth]{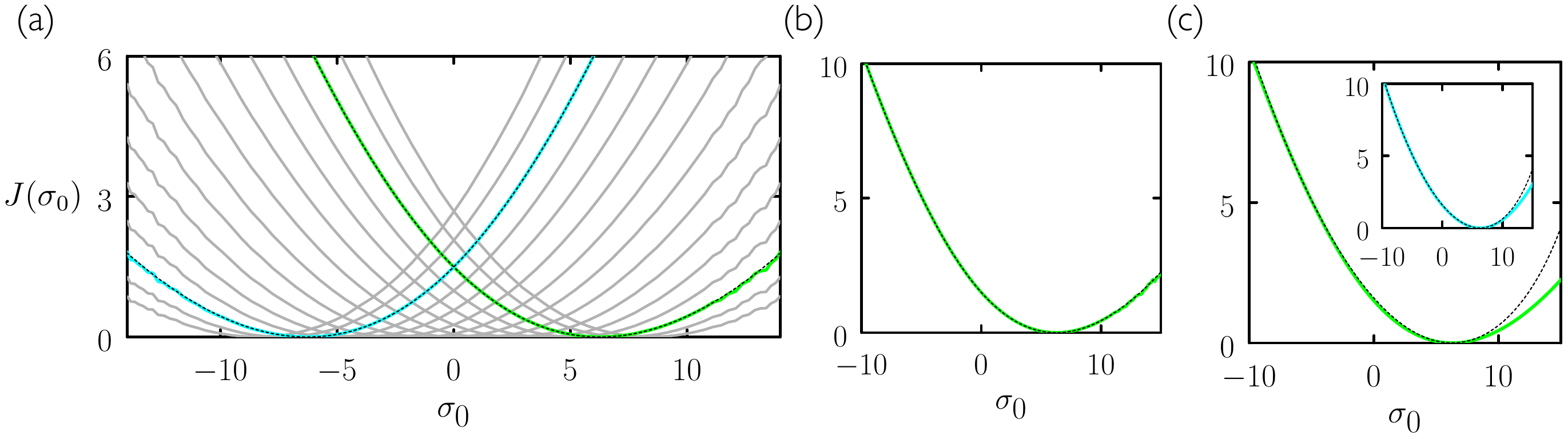} 
    \caption{As \f{fig1}, but for trajectories of constant time rather than constant event number; here $J(\sigma_0) = -T^{-1} \ln \rho(\sigma_0,T)$. (a) The black dashed lines, $J_0(\sigma_0)+\sigma_0$ on the left and $J_1(\sigma_0)-\sigma_0$ on the right, verify that the fluctuation relation \eq{fr} holds. (b) Numerically computed rate function for the model $s=0$ (green), versus the exact solution (black) calculated via matrix diagonalization. (c) Numerically computed rate-function upper bound, \eqq{three2} (black), versus the exact solution (green). Inset: bound from the present method (black), compared with the thermodynamic uncertainty relation from Ref.\c{gingrich2016dissipation} (cyan; this is the upper line from the main panel of Fig. 2 of that reference).}
    \label{fig_ct}
 \end{figure*}
 \\~
\indent These quantities allow the approximation, via \eqq{two}, of the large-deviation rate function $I_s(\sigma_0)=-K^{-1} \ln \rho_s(\sigma_0,K)$ for 15 members of the family \eq{rm}. Results are shown in \f{fig1}(a). Recall that the variable $\sigma_0$ is the entropy production rate defined with respect to the model $s=0$. The green and cyan curves correspond to the models $s=0$ and $s=1$ respectively, which are the forward and reverse versions of the model with rates $W_0(C \to C')$. The black dotted lines verify that these models obey the fluctuation relation \eq{oneb}. The left-hand black dotted line is $I_0(\sigma_0) +\sigma_0$, which is equal to $I_1(\sigma_0)$ (cyan line). The right-hand black dotted line is $I_1(\sigma_0) -\sigma_0$, which is equal to $I_0(\sigma_0)$ (green). This comparison provides a nontrivial check on the numerics, because it involves the superposition of the centers and tails of distinct distributions.  As a further check on our numerics we show in \f{fig1}(b) a comparison between the rate function calculated by the procedure described here, and that obtained by Legendre transform of the logarithm of the largest eigenvalue of the tilted transition-probability matrix\c{touchette2009large,chiuchiu2017mapping}. More generally, the distributions shown satisfy $I_s(\sigma_0)= I_{1-s}(-\sigma_0)$. 
\\~
\indent Recent work has identified rate-function bounds for generalized currents for discrete-time and continuous-time Markov chains\c{proesmans2017discrete,pietzonka2017finite,gingrich2016dissipation,horowitz2017proof,chiuchiu2017mapping}. The rate-function upper bound obtained from the present method, \eqq{three}, can be evaluated analytically for simple models or, in general, using a single reference-model trajectory\c{lattice2}. The bound is not necessarily quadratic or convex. Combined with the multi-point sampling method, the bound provides a cheap way of estimating the large-deviation rate function for a range of models and conditions. For the present model and order parameter, the bound is relatively tight: see \f{fig1}(c).   
\\~
\indent Here we have demonstrated that multiple rate functions for path-extensive observables of Markov chains of a fixed number of configuration changes can be calculated in a simple way using a set of stochastic simulations. The rate functions of more complex models, e.g. lattice models, can be obtained in a similar way\c{klymko2017rare,lattice1,lattice2}.
\\~
\section{Sampling trajectories of fixed time}
\label{cts}
\subsection{General considerations}
\label{cts_sub}
The procedure described thus far applies to Markov chains in which the total number of configuration changes $K$ is fixed. Here we present one way to adapt this method to trajectories of fixed time $T$. There are more sources of fluctuation in this ensemble, because upon making a move we select both a new microstate and a time increment, the latter being an exponentially-distributed random variable\c{gillespie2005general}. In general we must impose a bias on both choices in order to sample dynamic observables. 
 
Consider a set of models with rates
\beq
\label{rm2}
\hat{W}_{s}^{\lambda}(C \to C') = \Gamma_s^\lambda(C) W_s(C \to C').
\eeq
Here $W_s(C \to C')$ is given by \eq{rm}, and biases the choice of microstate $C'$. The factor $\Gamma_\lambda(C)$ is chosen to bias the jump time from microstate $C$. A convenient choice is to take 
\beq
\Gamma_s^{\lambda}(C) = \frac{R_0(C)+\lambda}{R_s(C)},
\eeq
which makes the escape rate of the new model to be a constant shift different to that of the original model, i.e. $\hat{R}_s^\lambda(C) \equiv \sum_{C'} \hat{W}_s^{\lambda}(C \to C') = \Gamma_s^\lambda(C) R_s(C) = R_0(C) + \lambda$. Here $\lambda >- \min_C R_0(C)$ is a constant that serves to make the escape times from microstate $C$ unusually large or small (by the reckoning of the original model). It has no effect upon the choice of microstate, i.e.
\beq
\label{move}
\hat{p}_s(C \to C') = \frac{\hat{W}_s^{\lambda}(C \to C')}{\hat{R}_s^\lambda(C)} = p_s(C \to C'),
\eeq
as \eqq{pb}.

We wish to calculate~\footnote{For simplicity we omit the hat on $\rho_s^{\lambda}(a,T)$.}
\beq
\label{pdf2}
\rho_s^\lambda(a,T) \equiv \sum_\x{\hat{P}_s^{\lambda}[\x] \delta{(A[\x]-T a)} \delta{(T[\x]-T)}},
\eeq
where now 
\beq
\sum_{\x}=\sum_{K>0}\sum_{C_0\cdots C_K} \prod_{k=0}^{K-1} \int_0^\infty {\rm d} t_k
\eeq
denotes a sum over trajectories $\x$ that can have variable numbers of configuration changes and jump times\c{budini2014fluctuating,giardina2011simulating}. [Using this notation, \eqq{pdf} would be written
\beq
\rho_s(a,K) \equiv \sum_\x{P_s[\x] \delta{(A[\x]-K a)}\delta{(K[\x]-K)}}.]\eeq
The quantity $T[\x]=\sum_{k=0}^{K[\x]-1} t_k$ is the total time of trajectory $\x$; $t_k$ is the time increment of configuration change $k+1$, and is $-\ln \eta/(R_0(C_k)+\lambda)$, where $\eta$ is a random number uniformly drawn from the interval $(0,1]$\c{gillespie2005general}.

The (bulk) path weight $\hat{P}_s^\lambda[\x]$ for the models $\hat{W}_s^\lambda$ is
\bea
\hat{P}_s^\lambda[\x] &=& \prod_{k=0}^{K[\x]-1} \hat{W}_s^{\lambda}(C_k \to C_{k+1}) \e^{-[R_0(C_k)+\lambda] t_k}.
\eea
The relative weight $\hat{w}_{ss'}^{\lambda \lambda'}[\x]=\hat{P}_s^\lambda[\x]/\hat{P}_{s'}^{\lambda'}[\x]$ is
\beq
\label{weight}
\hat{w}_{ss'}^{\lambda \lambda'} [\x] =e^{(s'-s) A[\x]+ (\lambda'-\lambda) T[\x]+ T \phi_{ss'}^{\lambda \lambda'}[\x]},
\eeq
where $A[\x] = \sum_{k=0}^{K[\x]-1} \alpha(C_k \to C_{k+1})$ is the extensive dynamic order parameter of trajectory $\x$, and
\beq
\label{kew2}
\phi_{ss'}^{\lambda \lambda'}[\x] \equiv T^{-1} \sum_{k=0}^{K[\x]-1} \ln \left(\frac{R_{s'}(C_k)}{R_s(C_k)} \frac{R_0(C)+\lambda}{R_0(C)+\lambda'}\right)
\eeq
is the analog of \eq{kew}, with $T$ replacing $K$ as the measure of path length. Note that for fixed $A[\x]=A$ and $T[\x]=T$ the path weight \eq{weight} depends, as before, only on configurations visited. It does not depend upon the time of jumps, on account of all models $\hat{W}_s^\lambda$ possessing identically shifted escape rates from a given configuration. Biasing jump times allows us to overcome the problem of sampling exponential tails of jump-time distributions using direct simulation (see Appendix B of Ref.\c{klymko2017rare}).
 
The quantity we want, the probability distribution of $a=A/K$ for trajectories of the original model of length $T$, is
\bea
\label{com2}
\rho_s^\lambda(a,T)&\equiv& \sum_\x{\hat{P}_{s'}^{\lambda'}[\x] \hat{w}_{ss'}^{\lambda \lambda'}[\x] \delta{(A[\x]-T a)} \delta{(T[\x]-T)}} \nonumber\\
&=& \rho_{s'}^{\lambda'}(a,T) \e^{\left[(s'-s) a+(\lambda'-\lambda)\right]T} \av{\e^{T \phi_{ss'}^{\lambda \lambda'}[\x]}}_{s'\lambda'}^{a T}, \hspace{0.5cm}
\eea
where
\beq
\av{\cdot }_{s'\lambda'}^{a T} \equiv  \frac{\sum_\x{\hat{P}_{s'}^{\lambda'}[\x] (\cdot) \delta{(A[\x]-T a)} \delta{(T[\x]-T)}}}{\sum_\x{\hat{P}_{s'}^{\lambda'}[\x] \delta{(A[\x]-T a)} \delta{(T[\x]-T)}}}
\eeq
is an average over reference-model trajectories of total time $T$ that possess $A[\x]= Ta$. \eqq{com2} can be written 
\bea
\label{fluc2}
\frac{\rho_s^{\lambda}(a ,T)}{\rho_{s'}^{\lambda'}(a ,T)} &=& \e^{\left[(s'-s) a+ (\lambda'-\lambda) \right]T} \nonumber \\
&\times& \int {\rm d} \phi_{ss'}^{\lambda \lambda'} \, \e^{T \phi_{ss'}^{\lambda \lambda'}} P_{s'}(\phi_{ss'}^{\lambda \lambda'}|a),
\eea
where $P_{s'}^{\lambda'}(\phi_{ss'}^{\lambda \lambda'}|a)$ is the probability distribution of the fluctuating piece of the path weight, $\phi_{ss'}^{\lambda \lambda'}[\x]$, for trajectories of the reference model $(s',\lambda')$ that possess $A[\x]=T a$. Normalization is such that $\int {\rm d} \phi_{ss'}^{\lambda \lambda'}\, P_{s'}(\phi_{ss'}^{\lambda \lambda'}|a)=1$. 

For models and observables for which, for long times, the large-deviation rate function $J_{s \lambda}(a) = -T^{-1} \ln \rho_s^\lambda(a ,T)$ exists, we can evaluate \eq{fluc2} at the concentration point $a=a_{s'\lambda'}$ of the reference model, where $J_{s'}^{\lambda'}(a_{s'\lambda'})=0$, to give
\bea
\label{one2}
J_{s \lambda}(a_{s'\lambda'}) &=& (s-s') a_{s'\lambda'} +(\lambda-\lambda') -\bar{\phi}_{ss'}^{\lambda \lambda'} \\
&-&T^{-1} \ln \int {\rm d} \phi_{ss'}^{\lambda \lambda'} \, \e^{T (\phi_{ss'}^{\lambda \lambda'}-\bar{\phi}_{ss'}^{\lambda \lambda'})} P_{s'}^{\lambda'}(\phi_{ss'}^{\lambda \lambda'}|a_{s'\lambda'}). \nonumber
\eea
Here $\bar{\phi}_{ss'}^{\lambda \lambda'}$ is the mean of the quantity $\phi_{ss'}^{\lambda \lambda'}[\x]$, for trajectories of the reference model $(s',\lambda')$ of length $T$ that have $A[\x] = T a_{s'\lambda'}$. By Jensen's inequality, the first line on the right-hand side of \eq{one2} provides an upper bound 
\beq
\label{three2}
J_{s\lambda}^{(0)}(a_{s'\lambda'}) = (s-s') a_{s'\lambda'}+ (\lambda-\lambda')-\bar{\phi}_{ss'}^{\lambda \lambda'}
\eeq
on the rate function. A refinement to this estimate can be computed by assuming $P_{s'}^{\lambda'}(\phi_{ss'}^{\lambda \lambda'}|a_{s'})$ to be Gaussian, in which case we get
\beq
\label{two2}
J^{(1)}_{s\lambda}(a_{s'\lambda'}) = J_{s\lambda}^{(0)}(a_{s'\lambda'})-T {\sigma^2}_{ss'}^{\lambda \lambda'}/2.
\eeq
Here ${\sigma^2}_{ss'}^{\lambda \lambda'}$ is the variance of the quantity $\phi_{ss'}^{\lambda \lambda'}[\x]$, for trajectories of the reference model $(s', \lambda')$ of length $T$ that have $A[\x] = T a_{s' \lambda'}$. Simulations of the reference model $(s',\lambda')$ for fixed time $T$ therefore provide an estimate of one point $a=a_{s'\lambda'}$ on the rate function $J_{s \lambda}(a)$.
 
 The procedure described in this section is similar to that described in \s{sampling}, but with path length set by time $T$ rather than number of configuration changes $K$. In both versions of the method the set of reference models is guided toward unlikely states of the original model with biased probabilities $p_s(C \to C)$. The additional consideration required for constant time is to draw random numbers $\eta$ in order to compute jump times $-\ln \eta/(R_0(C_k)+\lambda)$, using $\lambda$ as a way of sampling unlikely time changes. The choice of a constant shift $\lambda$ removes the need to account for fluctuating jump times in the re-weighting factor. The fluctuations of the remaining piece, $\phi_{ss'}^{\lambda \lambda'}[\x]$, \eqq{kew2}, derive from fluctuations of the empirical measure of the reference model and from the fact that the number of terms in the sum, $K[\x]$, varies from trajectory to trajectory. What is required is to measure the statistics of this quantity for trajectories of fixed length $T$ and fixed $A[\x] = T a$ (rather than for trajectories of fixed $K$ and fixed $A[\x] = K a$). 
 
 \subsection{Entropy production}
To test this method we calculated the large-deviation rate function for entropy production for the 4-state model of Ref.\c{gingrich2016dissipation}, for trajectories of constant time. In principle we should sample both rare states and rare jump times, but we found that the rate function for entropy production can be well approximated, over a large interval, by sampling only states. In this case we set $\lambda' =0$ and suppress $\lambda$-labels on quantities (the original model has $s=\lambda=0$). With the bias in \eq{rm2} given by \eq{alpha}, the intensive dynamical order parameter $a$ for a trajectory $\x$ is $\sigma_0[\x] = T^{-1} \sum_{k=0}^{K[\x]-1} \alpha(C_k \to C_{k+1})$, the rate of entropy production. As described by the equations above, the procedure used to sample fluctuations of this variable for paths of fixed $T$ is essentially identical to that used in \s{ent}, except that now we run reference-model trajectories for a fixed time $T=10^6$ rather than for a fixed number of events. We generated $5 \times 10^3$ reference-model trajectories for each value of $s'$ (100 different values between $-2$ and $2.5$). The values of $a$ of these trajectories fluctuate around a typical value $a_{s'}$. We calculated the mean $\bar{\phi}_{ss'}$ and variance ${\sigma}_{ss'}$ using a series of narrow windows at and either side of the typical value $a_{s'}$, and constructed a weighted average of the values of $\sigma^2_{ss'}$ calculated in these windows. This weighted average was used to compute the approximation \eq{two2}. In this way we use most of the reference-model trajectories generated. 

We show the results of this procedure in \f{fig_ct}. Panel (a) shows that trajectories of constant time of the forward ($s=0$) and time-reversed ($s=1$) models obey the fluctuation relation \eq{fr}, as expected [the passage from \eq{fluc2} to \eq{fr}, for $\lambda = \lambda'=0$, is the same as that from \eq{fluc} to \eq{fr}], where now $\Sigma = \sigma_0 T$. Noise is visible in the tails of the numerical results -- for constant $T$ there are more sources of fluctuation than for constant $K$ -- but the rate functions can be reconstructed with reasonable accuracy within the interval shown. In panel (b) we compare the numerically computed approximation of the rate function, \eqq{two2} (green), with the exact solution (black). The latter was calculated by Legendre transform of the largest eigenvalue of the tilted rate matrix\c{lebowitz1999gallavotti,garrahan2009first,chiuchiu2017mapping}. In panel (c) we compare the exact solution (green) with the rate-function upper bound, \eqq{three2} (black), which is less tight in general than its constant-$K$ analog [see \f{fig1}(c)] but still close to the exact answer over a considerable interval. In the inset of the figure we compare the bound to the thermodynamic uncertainty relation\c{gingrich2016dissipation}; the two bounds are not exactly the same, but are similar over a large interval. The thermodynamic uncertainty relation provides a `master' upper bound for all currents. By contrast, the present approach provides a bound specific to the chosen observable, including observables that are not currents (see e.g. \f{fig_events}), as well as a scheme for improving upon the bound.

In the far wings of the rate functions (not shown) we found that calculation of the variance $\sigma^2_{ss'}$ was unreliable; there, more trajectories are needed. In addition, we would expect, for extremely rare values of entropy production, that trajectories resulting from unusual jump times must be sampled (using the $\lambda$-bias) in order to accurately compute the rate function. One simple approach that could be used here is to carry out two-dimensional sampling in $s'$ (biasing states) and $\lambda$ (biasing jump times), similar to that done in Sec. V of\c{klymko2017rare}. Such sampling would produce a set of rate-function bounds at least as tight as the one shown in \f{fig_ct}(c), and those bounds could be improved by evaluation of the fluctuation terms. However, given the close correspondence between numerics and the exact answer, such sampling does not appear to be necessary for the range of values of $\sigma_0$ shown in the figure.

\subsection{Events}
 \begin{figure}[b] 
    \centering
    \includegraphics[width=0.9\linewidth]{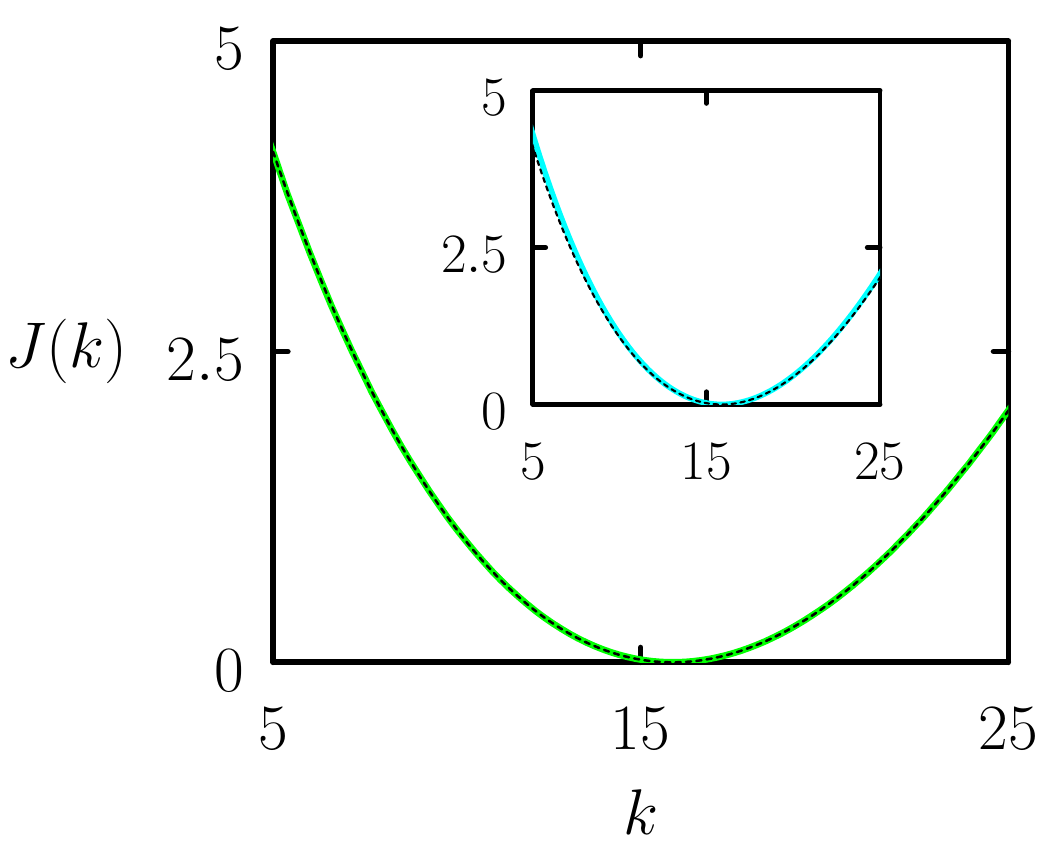} 
    \caption{Large-deviation rate function $J(k) =- T^{-1} \ln \rho(k,T)$  for the number of configuration changes $k=K/T$ per unit time in the 4-state model of Ref.\c{gingrich2016dissipation}. The black line is the exact result, from matrix diagonalization. The green line results from numerical evaluation of \eqq{two2}, using the sampling procedure described in this paper. Inset: the cyan line is the upper bound \eq{three2}, estimated using a single reference-model trajectory. This bound is close to the exact answer (black).}
    \label{fig_events}
 \end{figure}
 
In studies of glass-like models, an important observable is the number of configuration changes (events) $K$ occurring in time $T$\c{garrahan2009first}. In order to sample
\beq
\label{pdf3}
\rho_0^0(K,T) \equiv \sum_\x{\hat{P}_0^0[\x] \delta{(K[\x]-K)} \delta{(T[\x]-T)}},
\eeq
we used the umbrella procedure of \s{cts_sub}, choosing only to bias jump times (using $\lambda$) and not the choice of state. In this case the reference model $\hat{W}_{0}^\lambda(C \to C')$ selects states in the same way as the original model, with probabilities $p_0(C \to C')$, and selects jump times from exponential distributions with mean $R_0(C) + \lambda$. 
The path weight \eq{kew2} simplifies to 
\beq
\label{kew3}
\phi_{00}^{0 \lambda'}[\x] \equiv T^{-1} \sum_{k=0}^{K[\x]-1} \ln \frac{R_0(C)}{R_0(C)+\lambda'}.
\eeq
We carried out the usual procedure in order to estimate the large-deviation rate function for events $k=K/T$ in a fixed time for the 4-state model of Ref.\c{gingrich2016dissipation}. We used 70 reference models whose values of $\lambda'$ were evenly spaced between $\pm 10$. For each value of $\lambda'$ we generated $2 \times 10^4$ trajectories of length $T=10^5$, and calculated for each set the typical value $a_{0\lambda'} = k_{0\lambda'}$, and the values of $\bar{\phi}_{00}^{0\lambda'}$ and $\sigma_{00}^{0\lambda'}$ (as in the previous subsection we used a series of windows around the typical value $a_{0\lambda'}$). We used these quantities to calculate the Gaussian weight-fluctuation approximation \eq{two2} to $J(k) = -T^{-1} \ln \rho(K,T)$ [a single trajectory suffices to calculate the bound \eq{three2}]. Note that in those equations we have $s=s'=0$.

Numerical results are shown in \f{fig_events}, compared with the exact result. The latter was obtained by Legendre transform of the largest eigenvalue of the tilted rate matrix\c{lebowitz1999gallavotti,garrahan2009first,chiuchiu2017mapping}. Both the bound and its refinement are close to the exact answer. 

\section{Conclusions}
\label{conc}

We have described a simple method of umbrella sampling for Markov chains. The method uses a simple modification of a model $W_0(C \to C')$, namely \eq{rm} or \eq{rm2}, to estimate the large-deviation rate functions of a path-extensive observable for that model. In addition, with the same set of simulations one can reconstruct the large-deviation rate functions for any member of the family \eq{rm} or \eq{rm2}. The general relationship between probability distributions of members of this family is an extended fluctuation relation [Eqs. \eq{fluc} or \eq{fluc2}]. When the observable is chosen to be the rate of entropy production, members of this family include the forward and reverse version of a certain model, whose probability distributions are related by a simple fluctuation relation, \eqq{oneb}.  The extension to multiple dynamic observables is possible, following Section V of\c{klymko2017rare} (see also section 2.3 of\c{lebowitz1999gallavotti}). 
~\\
\indent The method described here applies to Markov chains of a fixed number of configuration changes or fixed time, with computation of the rate function in the latter case being (in general) numerically more demanding. However, in both cases it is straightforward to compute a process-specific upper bound on the rate function, \eq{three} or \eq{three2}. The bound derives from the typical weight of a simply-modified version of the original process, and can be computed (for a particular value of the order parameter) using a single dynamic trajectory. The bound is not necessarily quadratic about its origin or convex. Combined with the multi-point sampling procedure, computation of the bound provides a simple and numerically cheap way to identify special points, such as dynamic phase transitions\c{klymko2017rare,lattice2}, in a model's parameter space. The method also allows one to improve upon the bound. In the cases studied here this can be done with reasonable numerical effort, and results in estimates for large-deviation rate functions for path-extensive observables that agree with the answers obtained by matrix diagonalization.

\begin{acknowledgements}
I thank John Edison, Katherine Klymko, and Zden\v ek Preisler for discussions, and thank Todd Gingrich for discussions and for sharing data from Ref.\c{gingrich2016dissipation}. This work was performed at the Molecular Foundry, Lawrence Berkeley National Laboratory, supported by the Office of Science, Office of Basic Energy Sciences, of the U.S. Department of Energy under Contract No. DE-AC02--05CH11231. 
\end{acknowledgements}


\begin{thebibliography}{41}%
\makeatletter
\providecommand \@ifxundefined [1]{%
 \@ifx{#1\undefined}
}%
\providecommand \@ifnum [1]{%
 \ifnum #1\expandafter \@firstoftwo
 \else \expandafter \@secondoftwo
 \fi
}%
\providecommand \@ifx [1]{%
 \ifx #1\expandafter \@firstoftwo
 \else \expandafter \@secondoftwo
 \fi
}%
\providecommand \natexlab [1]{#1}%
\providecommand \enquote  [1]{``#1''}%
\providecommand \bibnamefont  [1]{#1}%
\providecommand \bibfnamefont [1]{#1}%
\providecommand \citenamefont [1]{#1}%
\providecommand \href@noop [0]{\@secondoftwo}%
\providecommand \href [0]{\begingroup \@sanitize@url \@href}%
\providecommand \@href[1]{\@@startlink{#1}\@@href}%
\providecommand \@@href[1]{\endgroup#1\@@endlink}%
\providecommand \@sanitize@url [0]{\catcode `\\12\catcode `\$12\catcode
  `\&12\catcode `\#12\catcode `\^12\catcode `\_12\catcode `\%12\relax}%
\providecommand \@@startlink[1]{}%
\providecommand \@@endlink[0]{}%
\providecommand \url  [0]{\begingroup\@sanitize@url \@url }%
\providecommand \@url [1]{\endgroup\@href {#1}{\urlprefix }}%
\providecommand \urlprefix  [0]{URL }%
\providecommand \Eprint [0]{\href }%
\providecommand \doibase [0]{http://dx.doi.org/}%
\providecommand \selectlanguage [0]{\@gobble}%
\providecommand \bibinfo  [0]{\@secondoftwo}%
\providecommand \bibfield  [0]{\@secondoftwo}%
\providecommand \translation [1]{[#1]}%
\providecommand \BibitemOpen [0]{}%
\providecommand \bibitemStop [0]{}%
\providecommand \bibitemNoStop [0]{.\EOS\space}%
\providecommand \EOS [0]{\spacefactor3000\relax}%
\providecommand \BibitemShut  [1]{\csname bibitem#1\endcsname}%
\let\auto@bib@innerbib\@empty
\bibitem [{\citenamefont {Frenkel}\ and\ \citenamefont
  {Smit}(2001)}]{frenkel2001understanding}%
  \BibitemOpen
  \bibfield  {author} {\bibinfo {author} {\bibfnamefont {D.}~\bibnamefont
  {Frenkel}}\ and\ \bibinfo {author} {\bibfnamefont {B.}~\bibnamefont {Smit}},\
  }\href@noop {} {\emph {\bibinfo {title} {Understanding molecular simulation:
  from algorithms to applications}}},\ Vol.~\bibinfo {volume} {1}\ (\bibinfo
  {publisher} {Academic Press},\ \bibinfo {year} {2001})\BibitemShut {NoStop}%
\bibitem [{\citenamefont {Torrie}\ and\ \citenamefont
  {Valleau}(1977)}]{torrie1977nonphysical}%
  \BibitemOpen
  \bibfield  {author} {\bibinfo {author} {\bibfnamefont {G.}~\bibnamefont
  {Torrie}}\ and\ \bibinfo {author} {\bibfnamefont {J.}~\bibnamefont
  {Valleau}},\ }\href@noop {} {\bibfield  {journal} {\bibinfo  {journal}
  {Journal of Computational Physics}\ }\textbf {\bibinfo {volume} {23}},\
  \bibinfo {pages} {187} (\bibinfo {year} {1977})}\BibitemShut {NoStop}%
\bibitem [{\citenamefont {Bucklew}(1990)}]{bucklew1990large}%
  \BibitemOpen
  \bibfield  {author} {\bibinfo {author} {\bibfnamefont {J.~A.}\ \bibnamefont
  {Bucklew}},\ }\href@noop {} {\emph {\bibinfo {title} {Large deviation
  techniques in decision, simulation, and estimation}}}\ (\bibinfo  {publisher}
  {Wiley New York},\ \bibinfo {year} {1990})\BibitemShut {NoStop}%
\bibitem [{\citenamefont {Touchette}(2009)}]{touchette2009large}%
  \BibitemOpen
  \bibfield  {author} {\bibinfo {author} {\bibfnamefont {H.}~\bibnamefont
  {Touchette}},\ }\href@noop {} {\bibfield  {journal} {\bibinfo  {journal}
  {Physics Reports}\ }\textbf {\bibinfo {volume} {478}},\ \bibinfo {pages} {1}
  (\bibinfo {year} {2009})}\BibitemShut {NoStop}%
\bibitem [{\citenamefont {Garrahan}\ \emph {et~al.}(2009)\citenamefont
  {Garrahan}, \citenamefont {Jack}, \citenamefont {Lecomte}, \citenamefont
  {Pitard}, \citenamefont {van Duijvendijk},\ and\ \citenamefont {van
  Wijland}}]{garrahan2009first}%
  \BibitemOpen
  \bibfield  {author} {\bibinfo {author} {\bibfnamefont {J.~P.}\ \bibnamefont
  {Garrahan}}, \bibinfo {author} {\bibfnamefont {R.~L.}\ \bibnamefont {Jack}},
  \bibinfo {author} {\bibfnamefont {V.}~\bibnamefont {Lecomte}}, \bibinfo
  {author} {\bibfnamefont {E.}~\bibnamefont {Pitard}}, \bibinfo {author}
  {\bibfnamefont {K.}~\bibnamefont {van Duijvendijk}}, \ and\ \bibinfo {author}
  {\bibfnamefont {F.}~\bibnamefont {van Wijland}},\ }\href@noop {} {\bibfield
  {journal} {\bibinfo  {journal} {Journal of Physics A: Mathematical and
  Theoretical}\ }\textbf {\bibinfo {volume} {42}},\ \bibinfo {pages} {075007}
  (\bibinfo {year} {2009})}\BibitemShut {NoStop}%
\bibitem [{\citenamefont {Giardina}\ \emph {et~al.}(2006)\citenamefont
  {Giardina}, \citenamefont {Kurchan},\ and\ \citenamefont
  {Peliti}}]{giardina2006direct}%
  \BibitemOpen
  \bibfield  {author} {\bibinfo {author} {\bibfnamefont {C.}~\bibnamefont
  {Giardina}}, \bibinfo {author} {\bibfnamefont {J.}~\bibnamefont {Kurchan}}, \
  and\ \bibinfo {author} {\bibfnamefont {L.}~\bibnamefont {Peliti}},\
  }\href@noop {} {\bibfield  {journal} {\bibinfo  {journal} {Physical Review
  Letters}\ }\textbf {\bibinfo {volume} {96}},\ \bibinfo {pages} {120603}
  (\bibinfo {year} {2006})}\BibitemShut {NoStop}%
\bibitem [{\citenamefont {Maes}\ and\ \citenamefont
  {Neto{\v{c}}n{\`y}}(2008)}]{maes2008canonical}%
  \BibitemOpen
  \bibfield  {author} {\bibinfo {author} {\bibfnamefont {C.}~\bibnamefont
  {Maes}}\ and\ \bibinfo {author} {\bibfnamefont {K.}~\bibnamefont
  {Neto{\v{c}}n{\`y}}},\ }\href@noop {} {\bibfield  {journal} {\bibinfo
  {journal} {EPL (Europhysics Letters)}\ }\textbf {\bibinfo {volume} {82}},\
  \bibinfo {pages} {30003} (\bibinfo {year} {2008})}\BibitemShut {NoStop}%
\bibitem [{\citenamefont {Giardina}\ \emph {et~al.}(2011)\citenamefont
  {Giardina}, \citenamefont {Kurchan}, \citenamefont {Lecomte},\ and\
  \citenamefont {Tailleur}}]{giardina2011simulating}%
  \BibitemOpen
  \bibfield  {author} {\bibinfo {author} {\bibfnamefont {C.}~\bibnamefont
  {Giardina}}, \bibinfo {author} {\bibfnamefont {J.}~\bibnamefont {Kurchan}},
  \bibinfo {author} {\bibfnamefont {V.}~\bibnamefont {Lecomte}}, \ and\
  \bibinfo {author} {\bibfnamefont {J.}~\bibnamefont {Tailleur}},\ }\href@noop
  {} {\bibfield  {journal} {\bibinfo  {journal} {Journal of Statistical
  Physics}\ }\textbf {\bibinfo {volume} {145}},\ \bibinfo {pages} {787}
  (\bibinfo {year} {2011})}\BibitemShut {NoStop}%
\bibitem [{\citenamefont {Lecomte}\ and\ \citenamefont
  {Tailleur}(2007)}]{lecomte2007numerical}%
  \BibitemOpen
  \bibfield  {author} {\bibinfo {author} {\bibfnamefont {V.}~\bibnamefont
  {Lecomte}}\ and\ \bibinfo {author} {\bibfnamefont {J.}~\bibnamefont
  {Tailleur}},\ }\href@noop {} {\bibfield  {journal} {\bibinfo  {journal}
  {Journal of Statistical Mechanics: Theory and Experiment}\ }\textbf {\bibinfo
  {volume} {2007}},\ \bibinfo {pages} {P03004} (\bibinfo {year}
  {2007})}\BibitemShut {NoStop}%
\bibitem [{\citenamefont {Nemoto}\ and\ \citenamefont
  {Sasa}(2014)}]{nemoto2014computation}%
  \BibitemOpen
  \bibfield  {author} {\bibinfo {author} {\bibfnamefont {T.}~\bibnamefont
  {Nemoto}}\ and\ \bibinfo {author} {\bibfnamefont {S.-i.}\ \bibnamefont
  {Sasa}},\ }\href@noop {} {\bibfield  {journal} {\bibinfo  {journal} {Physical
  Review Letters}\ }\textbf {\bibinfo {volume} {112}},\ \bibinfo {pages}
  {090602} (\bibinfo {year} {2014})}\BibitemShut {NoStop}%
\bibitem [{\citenamefont {Rohwer}\ \emph {et~al.}(2015)\citenamefont {Rohwer},
  \citenamefont {Angeletti},\ and\ \citenamefont
  {Touchette}}]{rohwer2015convergence}%
  \BibitemOpen
  \bibfield  {author} {\bibinfo {author} {\bibfnamefont {C.~M.}\ \bibnamefont
  {Rohwer}}, \bibinfo {author} {\bibfnamefont {F.}~\bibnamefont {Angeletti}}, \
  and\ \bibinfo {author} {\bibfnamefont {H.}~\bibnamefont {Touchette}},\
  }\href@noop {} {\bibfield  {journal} {\bibinfo  {journal} {Physical Review
  E}\ }\textbf {\bibinfo {volume} {92}},\ \bibinfo {pages} {052104} (\bibinfo
  {year} {2015})}\BibitemShut {NoStop}%
\bibitem [{\citenamefont {Nemoto}\ \emph {et~al.}(2016)\citenamefont {Nemoto},
  \citenamefont {Bouchet}, \citenamefont {Jack},\ and\ \citenamefont
  {Lecomte}}]{nemoto2016population}%
  \BibitemOpen
  \bibfield  {author} {\bibinfo {author} {\bibfnamefont {T.}~\bibnamefont
  {Nemoto}}, \bibinfo {author} {\bibfnamefont {F.}~\bibnamefont {Bouchet}},
  \bibinfo {author} {\bibfnamefont {R.~L.}\ \bibnamefont {Jack}}, \ and\
  \bibinfo {author} {\bibfnamefont {V.}~\bibnamefont {Lecomte}},\ }\href@noop
  {} {\bibfield  {journal} {\bibinfo  {journal} {Physical Review E}\ }\textbf
  {\bibinfo {volume} {93}},\ \bibinfo {pages} {062123} (\bibinfo {year}
  {2016})}\BibitemShut {NoStop}%
\bibitem [{\citenamefont {Ray}\ \emph {et~al.}(2017{\natexlab{a}})\citenamefont
  {Ray}, \citenamefont {Chan},\ and\ \citenamefont
  {Limmer}}]{ray2017importance}%
  \BibitemOpen
  \bibfield  {author} {\bibinfo {author} {\bibfnamefont {U.}~\bibnamefont
  {Ray}}, \bibinfo {author} {\bibfnamefont {G.~K.}\ \bibnamefont {Chan}}, \
  and\ \bibinfo {author} {\bibfnamefont {D.~T.}\ \bibnamefont {Limmer}},\
  }\href@noop {} {\bibfield  {journal} {\bibinfo  {journal} {arXiv preprint
  arXiv:1708.00459}\ } (\bibinfo {year} {2017}{\natexlab{a}})}\BibitemShut
  {NoStop}%
\bibitem [{\citenamefont {Oakes}\ \emph {et~al.}(2018)\citenamefont {Oakes},
  \citenamefont {Powell}, \citenamefont {Castelnovo}, \citenamefont
  {Lamacraft},\ and\ \citenamefont {Garrahan}}]{jpgdimer}%
  \BibitemOpen
  \bibfield  {author} {\bibinfo {author} {\bibfnamefont {T.}~\bibnamefont
  {Oakes}}, \bibinfo {author} {\bibfnamefont {S.}~\bibnamefont {Powell}},
  \bibinfo {author} {\bibfnamefont {C.}~\bibnamefont {Castelnovo}}, \bibinfo
  {author} {\bibfnamefont {A.}~\bibnamefont {Lamacraft}}, \ and\ \bibinfo
  {author} {\bibfnamefont {J.~P.}\ \bibnamefont {Garrahan}},\ }\href@noop {}
  {\bibfield  {journal} {\bibinfo  {journal} {arXiv preprint arXiv:1802.09576}\
  } (\bibinfo {year} {2018})}\BibitemShut {NoStop}%
\bibitem [{\citenamefont {Ray}\ \emph {et~al.}(2017{\natexlab{b}})\citenamefont
  {Ray}, \citenamefont {Chan},\ and\ \citenamefont {Limmer}}]{ray2017exact}%
  \BibitemOpen
  \bibfield  {author} {\bibinfo {author} {\bibfnamefont {U.}~\bibnamefont
  {Ray}}, \bibinfo {author} {\bibfnamefont {G.~K.}\ \bibnamefont {Chan}}, \
  and\ \bibinfo {author} {\bibfnamefont {D.~T.}\ \bibnamefont {Limmer}},\
  }\href@noop {} {\bibfield  {journal} {\bibinfo  {journal} {arXiv preprint
  arXiv:1708.09482}\ } (\bibinfo {year} {2017}{\natexlab{b}})}\BibitemShut
  {NoStop}%
\bibitem [{\citenamefont {Klymko}\ \emph {et~al.}(2018)\citenamefont {Klymko},
  \citenamefont {Geissler}, \citenamefont {Garrahan},\ and\ \citenamefont
  {Whitelam}}]{klymko2017rare}%
  \BibitemOpen
  \bibfield  {author} {\bibinfo {author} {\bibfnamefont {K.}~\bibnamefont
  {Klymko}}, \bibinfo {author} {\bibfnamefont {P.~L.}\ \bibnamefont
  {Geissler}}, \bibinfo {author} {\bibfnamefont {J.~P.}\ \bibnamefont
  {Garrahan}}, \ and\ \bibinfo {author} {\bibfnamefont {S.}~\bibnamefont
  {Whitelam}},\ }\href {\doibase 10.1103/PhysRevE.97.032123} {\bibfield
  {journal} {\bibinfo  {journal} {Phys. Rev. E}\ }\textbf {\bibinfo {volume}
  {97}},\ \bibinfo {pages} {032123} (\bibinfo {year} {2018})}\BibitemShut
  {NoStop}%
\bibitem [{\citenamefont {Whitelam}(2018)}]{lattice2}%
  \BibitemOpen
  \bibfield  {author} {\bibinfo {author} {\bibfnamefont {S.}~\bibnamefont
  {Whitelam}},\ }\href {\doibase 10.1103/PhysRevE.97.032122} {\bibfield
  {journal} {\bibinfo  {journal} {Phys. Rev. E}\ }\textbf {\bibinfo {volume}
  {97}},\ \bibinfo {pages} {032122} (\bibinfo {year} {2018})}\BibitemShut
  {NoStop}%
\bibitem [{\citenamefont {Chetrite}\ and\ \citenamefont
  {Touchette}(2013)}]{chetrite2013nonequilibrium}%
  \BibitemOpen
  \bibfield  {author} {\bibinfo {author} {\bibfnamefont {R.}~\bibnamefont
  {Chetrite}}\ and\ \bibinfo {author} {\bibfnamefont {H.}~\bibnamefont
  {Touchette}},\ }\href@noop {} {\bibfield  {journal} {\bibinfo  {journal}
  {Physical Review Letters}\ }\textbf {\bibinfo {volume} {111}},\ \bibinfo
  {pages} {120601} (\bibinfo {year} {2013})}\BibitemShut {NoStop}%
\bibitem [{\citenamefont {Jack}\ and\ \citenamefont
  {Sollich}(2015)}]{jack2015effective}%
  \BibitemOpen
  \bibfield  {author} {\bibinfo {author} {\bibfnamefont {R.~L.}\ \bibnamefont
  {Jack}}\ and\ \bibinfo {author} {\bibfnamefont {P.}~\bibnamefont {Sollich}},\
  }\href@noop {} {\bibfield  {journal} {\bibinfo  {journal} {The European
  Physical Journal Special Topics}\ }\textbf {\bibinfo {volume} {224}},\
  \bibinfo {pages} {2351} (\bibinfo {year} {2015})}\BibitemShut {NoStop}%
\bibitem [{\citenamefont {Chetrite}\ and\ \citenamefont
  {Touchette}(2015)}]{chetrite2015variational}%
  \BibitemOpen
  \bibfield  {author} {\bibinfo {author} {\bibfnamefont {R.}~\bibnamefont
  {Chetrite}}\ and\ \bibinfo {author} {\bibfnamefont {H.}~\bibnamefont
  {Touchette}},\ }\href@noop {} {\bibfield  {journal} {\bibinfo  {journal}
  {Journal of Statistical Mechanics: Theory and Experiment}\ }\textbf {\bibinfo
  {volume} {2015}},\ \bibinfo {pages} {P12001} (\bibinfo {year}
  {2015})}\BibitemShut {NoStop}%
\bibitem [{\citenamefont {Garrahan}(2016)}]{garrahan2016classical}%
  \BibitemOpen
  \bibfield  {author} {\bibinfo {author} {\bibfnamefont {J.~P.}\ \bibnamefont
  {Garrahan}},\ }\href@noop {} {\bibfield  {journal} {\bibinfo  {journal}
  {Journal of Statistical Mechanics: Theory and Experiment}\ }\textbf {\bibinfo
  {volume} {2016}},\ \bibinfo {pages} {073208} (\bibinfo {year}
  {2016})}\BibitemShut {NoStop}%
\bibitem [{\citenamefont {Ferr{\'e}}\ and\ \citenamefont
  {Touchette}(2018)}]{ferre2018adaptive}%
  \BibitemOpen
  \bibfield  {author} {\bibinfo {author} {\bibfnamefont {G.}~\bibnamefont
  {Ferr{\'e}}}\ and\ \bibinfo {author} {\bibfnamefont {H.}~\bibnamefont
  {Touchette}},\ }\href@noop {} {\bibfield  {journal} {\bibinfo  {journal}
  {arXiv preprint arXiv:1803.11117}\ } (\bibinfo {year} {2018})}\BibitemShut
  {NoStop}%
\bibitem [{\citenamefont {Varadhan}(2010)}]{varadhan2010large}%
  \BibitemOpen
  \bibfield  {author} {\bibinfo {author} {\bibfnamefont {S.~S.}\ \bibnamefont
  {Varadhan}},\ }in\ \href@noop {} {\emph {\bibinfo {booktitle} {Proceedings of
  the International Congress of Mathematicians 2010 (ICM 2010) (In 4 Volumes)
  Vol. I: Plenary Lectures and Ceremonies Vols. II--IV: Invited Lectures}}}\
  (\bibinfo {organization} {World Scientific},\ \bibinfo {year} {2010})\ pp.\
  \bibinfo {pages} {622--639}\BibitemShut {NoStop}%
\bibitem [{\citenamefont {Gallavotti}\ and\ \citenamefont
  {Cohen}(1995)}]{gallavotti1995dynamical}%
  \BibitemOpen
  \bibfield  {author} {\bibinfo {author} {\bibfnamefont {G.}~\bibnamefont
  {Gallavotti}}\ and\ \bibinfo {author} {\bibfnamefont {E.~G.~D.}\ \bibnamefont
  {Cohen}},\ }\href@noop {} {\bibfield  {journal} {\bibinfo  {journal}
  {Physical Review Letters}\ }\textbf {\bibinfo {volume} {74}},\ \bibinfo
  {pages} {2694} (\bibinfo {year} {1995})}\BibitemShut {NoStop}%
\bibitem [{\citenamefont {Kurchan}(1998)}]{kurchan1998fluctuation}%
  \BibitemOpen
  \bibfield  {author} {\bibinfo {author} {\bibfnamefont {J.}~\bibnamefont
  {Kurchan}},\ }\href@noop {} {\bibfield  {journal} {\bibinfo  {journal}
  {Journal of Physics A: Mathematical and General}\ }\textbf {\bibinfo {volume}
  {31}},\ \bibinfo {pages} {3719} (\bibinfo {year} {1998})}\BibitemShut
  {NoStop}%
\bibitem [{\citenamefont {Maes}(1999)}]{maes1999fluctuation}%
  \BibitemOpen
  \bibfield  {author} {\bibinfo {author} {\bibfnamefont {C.}~\bibnamefont
  {Maes}},\ }\href@noop {} {\bibfield  {journal} {\bibinfo  {journal} {Journal
  of Statistical Physics}\ }\textbf {\bibinfo {volume} {95}},\ \bibinfo {pages}
  {367} (\bibinfo {year} {1999})}\BibitemShut {NoStop}%
\bibitem [{\citenamefont {Lebowitz}\ and\ \citenamefont
  {Spohn}(1999)}]{lebowitz1999gallavotti}%
  \BibitemOpen
  \bibfield  {author} {\bibinfo {author} {\bibfnamefont {J.~L.}\ \bibnamefont
  {Lebowitz}}\ and\ \bibinfo {author} {\bibfnamefont {H.}~\bibnamefont
  {Spohn}},\ }\href@noop {} {\bibfield  {journal} {\bibinfo  {journal} {Journal
  of Statistical Physics}\ }\textbf {\bibinfo {volume} {95}},\ \bibinfo {pages}
  {333} (\bibinfo {year} {1999})}\BibitemShut {NoStop}%
\bibitem [{\citenamefont {Crooks}(2000)}]{crooks2000path}%
  \BibitemOpen
  \bibfield  {author} {\bibinfo {author} {\bibfnamefont {G.~E.}\ \bibnamefont
  {Crooks}},\ }\href@noop {} {\bibfield  {journal} {\bibinfo  {journal}
  {Physical Review E}\ }\textbf {\bibinfo {volume} {61}},\ \bibinfo {pages}
  {2361} (\bibinfo {year} {2000})}\BibitemShut {NoStop}%
\bibitem [{\citenamefont {Seifert}(2005)}]{seifert2005entropy}%
  \BibitemOpen
  \bibfield  {author} {\bibinfo {author} {\bibfnamefont {U.}~\bibnamefont
  {Seifert}},\ }\href@noop {} {\bibfield  {journal} {\bibinfo  {journal}
  {Physical Review Letters}\ }\textbf {\bibinfo {volume} {95}},\ \bibinfo
  {pages} {040602} (\bibinfo {year} {2005})}\BibitemShut {NoStop}%
\bibitem [{\citenamefont {Speck}\ \emph {et~al.}(2012)\citenamefont {Speck},
  \citenamefont {Engel},\ and\ \citenamefont {Seifert}}]{speck2012large}%
  \BibitemOpen
  \bibfield  {author} {\bibinfo {author} {\bibfnamefont {T.}~\bibnamefont
  {Speck}}, \bibinfo {author} {\bibfnamefont {A.}~\bibnamefont {Engel}}, \ and\
  \bibinfo {author} {\bibfnamefont {U.}~\bibnamefont {Seifert}},\ }\href@noop
  {} {\bibfield  {journal} {\bibinfo  {journal} {Journal of Statistical
  Mechanics: Theory and Experiment}\ }\textbf {\bibinfo {volume} {2012}},\
  \bibinfo {pages} {P12001} (\bibinfo {year} {2012})}\BibitemShut {NoStop}%
\bibitem [{\citenamefont {Dembo}\ and\ \citenamefont
  {Zeitouni}(2010)}]{dembo2010large}%
  \BibitemOpen
  \bibfield  {author} {\bibinfo {author} {\bibfnamefont {A.}~\bibnamefont
  {Dembo}}\ and\ \bibinfo {author} {\bibfnamefont {O.}~\bibnamefont
  {Zeitouni}},\ }\href@noop {} {\emph {\bibinfo {title} {Large deviations
  techniques and applications, volume 38 of Stochastic Modelling and Applied
  Probability}}}\ (\bibinfo  {publisher} {Springer-Verlag, Berlin},\ \bibinfo
  {year} {2010})\BibitemShut {NoStop}%
\bibitem [{\citenamefont {Pietzonka}\ \emph {et~al.}(2016)\citenamefont
  {Pietzonka}, \citenamefont {Barato},\ and\ \citenamefont
  {Seifert}}]{pietzonka2016universal}%
  \BibitemOpen
  \bibfield  {author} {\bibinfo {author} {\bibfnamefont {P.}~\bibnamefont
  {Pietzonka}}, \bibinfo {author} {\bibfnamefont {A.~C.}\ \bibnamefont
  {Barato}}, \ and\ \bibinfo {author} {\bibfnamefont {U.}~\bibnamefont
  {Seifert}},\ }\href@noop {} {\bibfield  {journal} {\bibinfo  {journal}
  {Physical Review E}\ }\textbf {\bibinfo {volume} {93}},\ \bibinfo {pages}
  {052145} (\bibinfo {year} {2016})}\BibitemShut {NoStop}%
\bibitem [{\citenamefont {Gingrich}\ \emph {et~al.}(2016)\citenamefont
  {Gingrich}, \citenamefont {Horowitz}, \citenamefont {Perunov},\ and\
  \citenamefont {England}}]{gingrich2016dissipation}%
  \BibitemOpen
  \bibfield  {author} {\bibinfo {author} {\bibfnamefont {T.~R.}\ \bibnamefont
  {Gingrich}}, \bibinfo {author} {\bibfnamefont {J.~M.}\ \bibnamefont
  {Horowitz}}, \bibinfo {author} {\bibfnamefont {N.}~\bibnamefont {Perunov}}, \
  and\ \bibinfo {author} {\bibfnamefont {J.~L.}\ \bibnamefont {England}},\
  }\href@noop {} {\bibfield  {journal} {\bibinfo  {journal} {Physical Review
  Letters}\ }\textbf {\bibinfo {volume} {116}},\ \bibinfo {pages} {120601}
  (\bibinfo {year} {2016})}\BibitemShut {NoStop}%
\bibitem [{\citenamefont {Whitelam}\ \emph {et~al.}(2018)\citenamefont
  {Whitelam}, \citenamefont {Klymko},\ and\ \citenamefont {Mandal}}]{lattice1}%
  \BibitemOpen
  \bibfield  {author} {\bibinfo {author} {\bibfnamefont {S.}~\bibnamefont
  {Whitelam}}, \bibinfo {author} {\bibfnamefont {K.}~\bibnamefont {Klymko}}, \
  and\ \bibinfo {author} {\bibfnamefont {D.}~\bibnamefont {Mandal}},\
  }\href@noop {} {\bibfield  {journal} {\bibinfo  {journal} {The Journal of
  Chemical Physics}\ }\textbf {\bibinfo {volume} {148}},\ \bibinfo {pages}
  {154902} (\bibinfo {year} {2018})}\BibitemShut {NoStop}%
\bibitem [{\citenamefont {Chiuchi\`u}\ and\ \citenamefont
  {Pigolotti}(2018)}]{chiuchiu2017mapping}%
  \BibitemOpen
  \bibfield  {author} {\bibinfo {author} {\bibfnamefont {D.}~\bibnamefont
  {Chiuchi\`u}}\ and\ \bibinfo {author} {\bibfnamefont {S.}~\bibnamefont
  {Pigolotti}},\ }\href {\doibase 10.1103/PhysRevE.97.032109} {\bibfield
  {journal} {\bibinfo  {journal} {Phys. Rev. E}\ }\textbf {\bibinfo {volume}
  {97}},\ \bibinfo {pages} {032109} (\bibinfo {year} {2018})}\BibitemShut
  {NoStop}%
\bibitem [{\citenamefont {Proesmans}\ and\ \citenamefont {Van~den
  Broeck}(2017)}]{proesmans2017discrete}%
  \BibitemOpen
  \bibfield  {author} {\bibinfo {author} {\bibfnamefont {K.}~\bibnamefont
  {Proesmans}}\ and\ \bibinfo {author} {\bibfnamefont {C.}~\bibnamefont
  {Van~den Broeck}},\ }\href@noop {} {\bibfield  {journal} {\bibinfo  {journal}
  {EPL (Europhysics Letters)}\ }\textbf {\bibinfo {volume} {119}},\ \bibinfo
  {pages} {20001} (\bibinfo {year} {2017})}\BibitemShut {NoStop}%
\bibitem [{\citenamefont {Pietzonka}\ \emph {et~al.}(2017)\citenamefont
  {Pietzonka}, \citenamefont {Ritort},\ and\ \citenamefont
  {Seifert}}]{pietzonka2017finite}%
  \BibitemOpen
  \bibfield  {author} {\bibinfo {author} {\bibfnamefont {P.}~\bibnamefont
  {Pietzonka}}, \bibinfo {author} {\bibfnamefont {F.}~\bibnamefont {Ritort}}, \
  and\ \bibinfo {author} {\bibfnamefont {U.}~\bibnamefont {Seifert}},\
  }\href@noop {} {\bibfield  {journal} {\bibinfo  {journal} {Physical Review
  E}\ }\textbf {\bibinfo {volume} {96}},\ \bibinfo {pages} {012101} (\bibinfo
  {year} {2017})}\BibitemShut {NoStop}%
\bibitem [{\citenamefont {Horowitz}\ and\ \citenamefont
  {Gingrich}(2017)}]{horowitz2017proof}%
  \BibitemOpen
  \bibfield  {author} {\bibinfo {author} {\bibfnamefont {J.~M.}\ \bibnamefont
  {Horowitz}}\ and\ \bibinfo {author} {\bibfnamefont {T.~R.}\ \bibnamefont
  {Gingrich}},\ }\href@noop {} {\bibfield  {journal} {\bibinfo  {journal}
  {Physical Review E}\ }\textbf {\bibinfo {volume} {96}},\ \bibinfo {pages}
  {020103} (\bibinfo {year} {2017})}\BibitemShut {NoStop}%
\bibitem [{\citenamefont {Gillespie}(2005)}]{gillespie2005general}%
  \BibitemOpen
  \bibfield  {author} {\bibinfo {author} {\bibfnamefont {D.}~\bibnamefont
  {Gillespie}},\ }\href@noop {} {\bibfield  {journal} {\bibinfo  {journal}
  {Journal of Computational Physics}\ }\textbf {\bibinfo {volume} {22}},\
  \bibinfo {pages} {403} (\bibinfo {year} {2005})}\BibitemShut {NoStop}%
\bibitem [{Note1()}]{Note1}%
  \BibitemOpen
  \bibinfo {note} {For simplicity we omit the hat on $\rho _s^{\lambda
  }(a,T)$.}\BibitemShut {Stop}%
\bibitem [{\citenamefont {Budini}\ \emph {et~al.}(2014)\citenamefont {Budini},
  \citenamefont {Turner},\ and\ \citenamefont
  {Garrahan}}]{budini2014fluctuating}%
  \BibitemOpen
  \bibfield  {author} {\bibinfo {author} {\bibfnamefont {A.~A.}\ \bibnamefont
  {Budini}}, \bibinfo {author} {\bibfnamefont {R.~M.}\ \bibnamefont {Turner}},
  \ and\ \bibinfo {author} {\bibfnamefont {J.~P.}\ \bibnamefont {Garrahan}},\
  }\href@noop {} {\bibfield  {journal} {\bibinfo  {journal} {Journal of
  Statistical Mechanics: Theory and Experiment}\ }\textbf {\bibinfo {volume}
  {2014}},\ \bibinfo {pages} {P03012} (\bibinfo {year} {2014})}\BibitemShut
  {NoStop}%
\end{thebibliography}

%

\end{document}